\documentclass[nofootinbib, twocolumn, longbibliography,superscriptaddress]{revtex4-2}
\usepackage{orcidlink}
\usepackage{amsmath}
\usepackage{amsthm}
\usepackage{amssymb}
\usepackage{graphicx}
\usepackage{hyperref}
\usepackage{color}
\usepackage{array}
\usepackage[makeroom]{cancel}
\usepackage[capitalize]{cleveref}
\usepackage{changes}
\usepackage{qcircuit}
\usepackage{braket}
\usepackage{bbm}
\usepackage{bm}
\usepackage{booktabs}
\usepackage{titlesec}
\usepackage{comment}
\usepackage{subfigure}
\usepackage{algorithm}
\usepackage{algpseudocode}
\usepackage{adjustbox}
\DeclareMathOperator{\id}{\mathbbm{1}}

\newcommand{\beq}{\begin{equation}}
\newcommand{\eeq}{\end{equation}}
\newcommand{\PREP}{\mathrm{PREP}}
\newcommand{\SEL}{\mathrm{SEL}}
\newcommand{\QPE}{\text{QPE}}
\usepackage{multirow}
\usepackage{cellspace} 
\definecolor{MyDarkBlue}{rgb}{0,0.1,0.75}
\hypersetup{pdfborder={0 0 0},colorlinks,breaklinks=true,
  urlcolor={MyDarkBlue},citecolor={MyDarkBlue},linkcolor={MyDarkBlue}
}

\def\ket#1{\left| #1 \right\rangle}
\def\bra#1{\left\langle #1 \right|}

\theoremstyle{plain}

\theoremstyle{definition}

\theoremstyle{remark}

\usepackage{empheq}

\newcommand{\ba}{\begin{align}}
\newcommand{\ea}{\end{align}}

\newcommand{\bD}{\bm{D}}

\newcommand{\eff}{\text{eff}}

\newcommand{\tr}{\text{tr}}

\crefname{thm}{Theorem}{Theorems}
\crefname{dfn}{Definition}{Definitions}
\crefname{rmk}{Remark}{Remarks}
\crefname{lem}{Lemma}{Lemmas}
\crefname{cor}{Corollary}{Corollaries}

\newcommand{\error}{\varepsilon}

\renewcommand{\theparagraph}{\arabic{paragraph}.}

\titleformat{\paragraph}[runin]
  {\normalfont\itshape}
  {\theparagraph}
  {0.4em}
  {}
  
 \titlespacing{\paragraph}
  {10pt}{0.2\baselineskip}{0.3\baselineskip}

\begin{document}
\title{Simulating optically-active spin defects with a quantum computer}

\author{Jack S. Baker \orcidlink{https://orcid.org/0000-0001-6635-1397}}
\affiliation{Xanadu, Toronto, ON, M5G 2C8, Canada}

\author{Pablo A. M. Casares \orcidlink{https://orcid.org/0000-0001-5500-9115}}
\affiliation{Xanadu, Toronto, ON, M5G 2C8, Canada}

\author{Modjtaba Shokrian Zini}
\affiliation{Xanadu, Toronto, ON, M5G 2C8, Canada}

\author{Jaydeep Thik}
\affiliation{Toyota Research Institute of North America, Ann Arbor, MI, 48105, USA}

\author{Debasish Banerjee \orcidlink{https://orcid.org/https://orcid.org/0000-0003-2402-2944}}
\affiliation{Toyota Research Institute of North America, Ann Arbor, MI, 48105, USA}

\author{Chen Ling \orcidlink{https://orcid.org/https://orcid.org/0000-0002-8667-9792}}
\email{chen.ling@toyota.com}
\affiliation{Toyota Research Institute of North America, Ann Arbor, MI, 48105, USA}

\author{Alain Delgado \orcidlink{https://orcid.org/https://orcid.org/0000-0003-4225-6904}}
\email{alaindelgado@xanadu.ai}
\affiliation{Xanadu, Toronto, ON, M5G 2C8, Canada}

\author{Juan Miguel Arrazola \orcidlink{https://orcid.org/0000-0002-0619-9650}}
\affiliation{Xanadu, Toronto, ON, M5G 2C8, Canada}

\begin{abstract}
There is a pressing need for more accurate computational simulations of the opto-electronic properties of defects in materials to aid in the development of quantum sensing platforms. In this work, we explore how quantum computers could be effectively utilized for this purpose. Specifically, we develop fault-tolerant quantum algorithms to simulate optically active defect states and their radiative emission rates. We employ quantum defect embedding theory to translate the Hamiltonian of a defect-containing supercell into a smaller, effective Hamiltonian that accounts for dielectric screening effects. Our approach integrates block-encoding of the dipole operator with quantum phase estimation to selectively sample the optically active excited states that exhibit the largest dipole transition amplitudes. We also provide estimates of the quantum resources required to simulate a negatively-charged boron vacancy in a hexagonal boron nitride cluster. We conclude by offering a forward-looking perspective on the potential of quantum computers to enhance quantum sensor capabilities and identify specific scenarios where quantum computing can resolve problems traditionally challenging for classical computers.
\end{abstract}
\maketitle

\section{Introduction}\label{sec:intro}
Quantum defects in solid materials are promising platforms to develop quantum technologies including solid-state qubits, single-photon emitters, and quantum sensors~\cite{awschalom2013quantum, bassett2019quantum,atature2018material}. Well-known examples of quantum defects include the nitrogen- and silicon-vacancy in diamond, and also divacancies and transition metal impurities in silicon carbide~\cite{schirhagl2014nitrogen, atature2018material}. More recently, the possibility of creating point defects in two-dimensional materials with narrow photoluminescence lines and weak phonon sidebands has opened the possibility to design quantum sensors for detecting small electric and magnetic fields~\cite{liu2022spin, stern2022room}. However, a key condition to unlock these applications is the existence of robust high-spin defect states that can be optically controlled. These are referred to as spin-active quantum defects.

Experiments to discover spin-active defects are challenging as they are influenced by their distribution and inhomogeneities in the sample material~\cite{stern2022room}. A key quantity to identify a spin-active defect is the optically-detected magnetic resonance (ODMR) contrast, which indicates the variation in the intensity of the photoluminescence (PL) peak as a function of the frequency of an external microwave field~\cite{mathur2022excited}. The position of the prominent features in the PL spectrum is determined by the energy of the lowest-lying excited states of the defect, and the frequency at which the contrast signal is detected (ODMR frequency) corresponds to the energy spacing between the spin sublevels of the defect states. While the energies of the excited states are of the order of a few electron volts, the energy difference between the spin sublevels can be five to six orders of magnitude smaller \cite{gottscholl2020initialization, ivady2020ab}.
Consequently, ODMR experiments combine laser and microwave pulses to access the defect excited states while pumping excitations between the spin sublevels to measure the ODMR spectrum~\cite{stern2022room, gottscholl2021room}.

\begin{figure*}
    \centering
    \includegraphics[width=0.91\linewidth]{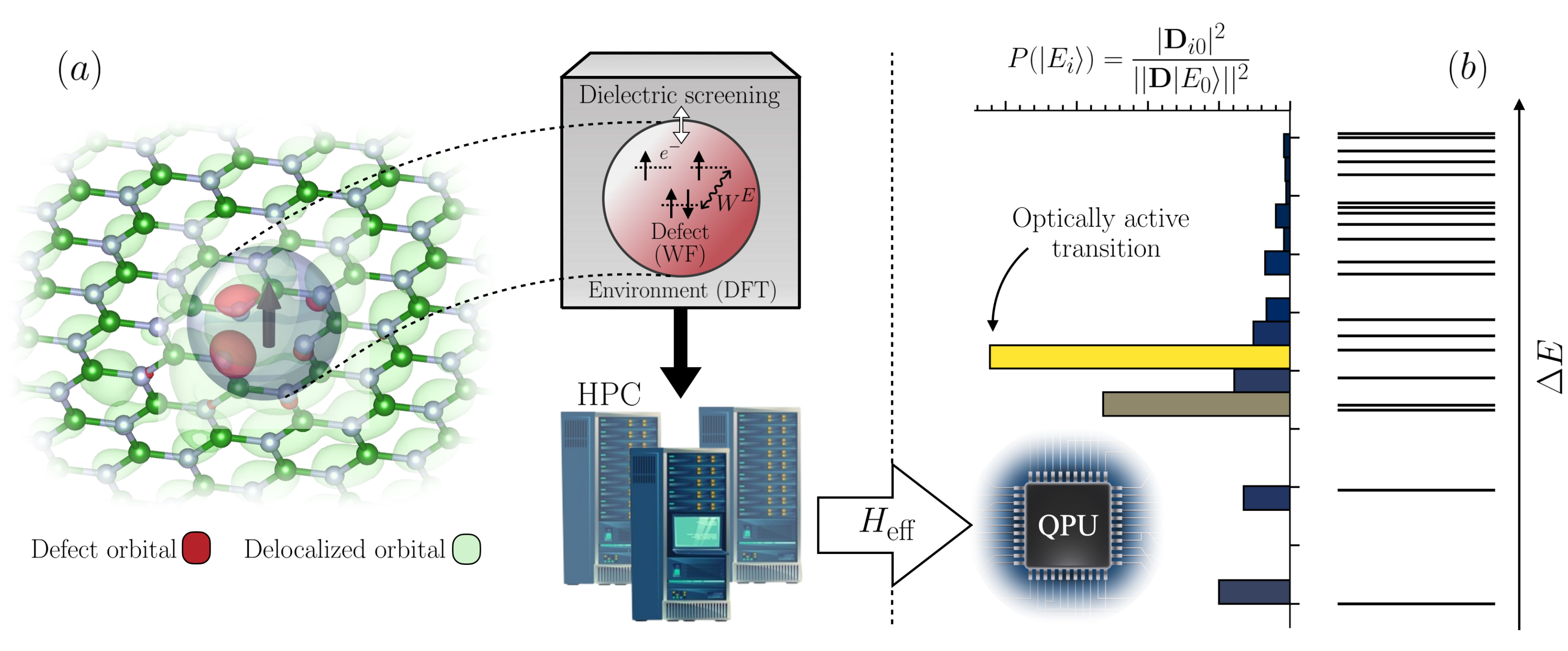}
    \caption{Overview of our approach combining quantum defect embedding theory (QDET) with a fault-tolerant quantum processing unit (QPU) to simulate the optically-active excited states of a spin defect in a material.  (a) The effective Hamiltonian $H_{\text{eff}}$ is constructed using simulations on a classical High Performance Computer (HPC) by considering the screened Coulomb interactions between electrons occupying local defect orbitals. This process is described in Section \ref{ssec:hamilt} and demonstrated practically in Section \ref{sec:application}. (b) The energy levels $E_i$, eigenstates $|E_i\rangle$ and ground-to-excited-state transition amplitudes $|\mathbf{D}_{i0}|^2$ are evaluated using the quantum algorithms described in Section \ref{sec:qlago}. Optically active excited states are sampled from the QPU with a probability proportional to $|\mathbf{D}_{i0}|^2$.} 
    \label{fig:hero}
\end{figure*}

Simulating accurately the low-lying excited states of defects is crucial for their optical characterization, and to simulate the fine structure of the defect states. For example, the components of the zero-field splitting (ZFS) tensor depend on the electronic structure of the defect states~\cite{ivady2018first}. While the ZFS tensor can be computed for the defect ground state using density functional theory (DFT), it is generally unknown for the excited states~\cite{mathur2022excited}. Furthermore, accurate calculations of the radiative decay and inter-system crossing rates between the defect states, which is key to investigate the ODMR activeness, must accurately capture electronic correlations at the level of excited states~\cite{bassett2019quantum}.

The required large size of the defect-containing supercells makes the application of highly-accurate classical wave function methods (e.g., full configuration interaction, multireference perturbation theory)~\cite{jensen2017} prohibitively expensive. Recently, the density-matrix renormalization group method was stretched to the limit to simulate a negative boron vacancy in hexagonal boron nitride (hBN) by considering an active space consisting of 27 Kohn-Sham orbitals out of the several hundred supercell bands~\cite{ivady2020ab}. An important drawback of this approach is that it neglects the screening effects due to the host material on the electronic structure of the quantum defect. On the other hand, time-dependent density functional theory (TDDFT) approximations can handle supercells with thousands of atoms~\cite{west-tddft}. However, TDDFT does not give access to the excited states, and its accuracy is limited by the approximate density functional and the adiabatic approximation~\cite{lacombe2020developing}. More accurate calculations of the excited states can be done by using the $GW$ approximation and the Bethe–Salpeter equation~\cite{gw-bse}, but this method is computationally expensive, and still affected by the underlying DFT approximation~\cite{qiu2013optical}. \par

Quantum embedding methods~\cite{sun2016quantum, ma2021quantum, qdet-paper, vorwerk2022quantum, wouters2016practicalDMET, dmft, muechler2022quantum} offer an avenue for simulating quantum defects in materials using wavefunction-based methods that are appealing to quantum computing. Crucially, they allow us to construct an effective Hamiltonian described using many less orbitals than the entire periodic supercell. Given the long runtimes for fault tolerant quantum simulations of solids with even a small number of atoms in the unit cell \cite{Ivanov2023}, this step considerably reduces the computational burden while retaining high accuracy. This is achieved by partitioning the system into a small active region describing the defect, and the environment consisting of the host material. In this framework, the electronic structure of the defect can be treated using a many-body approach run on a quantum computer, while accounting for the dielectric screening generated by the environment using lower-level classical simulations such as Hartree-Fock or density functional theory (DFT).

The development of fault-tolerant quantum algorithms to simulate the excited states and optical properties of quantum systems is a nascent area of research. Prior work has computed the Green’s function or polarizability tensor, which indirectly involves the dipole transition amplitudes~\cite{cai2020quantum, tong2021fast}. However, these references only discuss the straightforward application of available tools to these computational problems, without any resource estimation or optimization, raising questions about their practicality and the improvements required to implement them on quantum computers. Previous work has also proposed quantum algorithms to compute expectation values of arbitrary observables other than the Hamiltonian~\cite{steudtner2023fault}.
Other related works include variational approaches to these problems~\cite{kumar2023quantum,huang2022variational}, which are not guaranteed to succeed to the desired accuracy and suffer from scalability issues.

In this work, we develop quantum algorithms to estimate the energies of the optically-active excited states of the defect in different spin sectors with the largest dipole transition amplitudes. Our algorithms utilize optimized circuits for the block-encoding of the electric dipole operator leading to lower costs. Further, we show how to efficiently perform an amplitude estimation step that scales linearly, instead of quadratically, with the error, but requires a longer circuit depth and several controlled operations. We demonstrate how to control the involved subroutines only when necessary, saving gates and qubits. Thanks to the detailed compilation and constant factor resource estimation, we are able to accurately estimate the number of gates and qubits used by our algorithms to simulate the excited states and compute their energies and dipole transition amplitudes.

The paper is structured as follows.~\cref{sec:theory} describes the basics of the quantum embedding theory used to build the Hamiltonian of the quantum defect, and the observables required by the quantum algorithms. In~\cref{sec:qlago} we present the developed quantum algorithms. In \cref{sec:application} we validate the embedding approach to build the Hamiltonian for a negatively-charged boron vacancy in an hBN cluster. The contributions of Section \ref{sec:theory}-\ref{sec:application} are summarized in Fig. \ref{fig:hero}. In~\cref{sec:resources} we perform constant factor resource estimation for the developed algorithms. Finally,~\cref{sec:conclusions} summarizes the main conclusions.

\section{Theory}\label{sec:theory}
A promising quantum defect for sensing applications should exhibit the following key properties \cite{Koehl2015, bassett2019quantum, WolfowiczDefectByDesign}:
\begin{enumerate}
\item A strong and narrow optical emission line.
\item An ODMR spectrum with distinctive features at frequencies that are correlated with the magnitude of an external perturbation (e.g., magnetic field) to calibrate the quantum sensor.
\end{enumerate}

Full simulation of the ODMR spectrum is challenging as it requires computing the rates of radiative and non-radiative transitions between the defect states~\cite{ivady2020ab, ping2021-nature-simulations, bassett2019quantum}. Here, we focus on the simulation of the excited states for the optical characterization of quantum defects. We propose quantum algorithms to calculate the energies of the low-lying brightest excited states determining the position of the main peaks in the absorption or emission spectrum. To that aim we sample the eigenstates of the defect Hamiltonian with the largest dipole transition amplitudes, which can be used to estimate the radiative decay rates and lifetimes of the excited states. 

To enable quantum simulations of the opto-electronic properties of spin defects, it is crucial to work with a Hamiltonian whose size is significantly reduced compared with the Hamiltonian of the entire defect-containing supercell. Here we use the quantum defect embedding theory (QDET)~\cite{sheng2022green} to construct a significantly smaller (effective) Hamiltonian that can be simulated in a quantum computer with modest resources. We summarize the basics of the QDET methodology in the next section, and define the dipole moment operator used by the quantum algorithm in~\cref{sec:qlago} to sample the optically-active defect states.

\subsection{The effective Hamiltonian of the quantum defect}\label{ssec:hamilt}
The QDET method starts with a DFT simulation of the supercell to compute the energy bands at the $\Gamma$ point by using an approximate exchange-correlation density functional. The optimized Kohn-Sham (KS) states $\phi_n(\bm{r})$ are used to compute the localization factor~\cite{sheng2022green}
\begin{equation}
L_n = \int_{\text{V}_\text{D}} |\phi_n(\bm{r})|^2 d \bm{r},
\label{eq:loc_factor}
\end{equation}
to identify the single-particle states localized within the defect region with volume $\text{V}_\text{D}$. For a threshold value of $L_n$, we can select $N$ localized states with the largest localization factor to represent the many-electron wave function of the quantum defect. The number of electrons populating the selected defect states is determined by the occupation of the Kohn-Sham orbitals.

Then, the QDET method is applied to build the effective Hamiltonian describing the interacting electrons in the quantum defect. The dielectric screening due to the surrounding material is included at the level of the $G_0W_0$ approximation~\cite{sheng2022green}. The output of this calculation is the second-quantized electronic Hamiltonian given by
\begin{align}
H_\text{eff} &= \sum_{p,q=1}^N \sum_\sigma t_{pq}^\text{eff}  a_{p\sigma}^\dagger a_{q\sigma} \nonumber \\
& + \frac{1}{2} \sum_{p,q,r,s=1}^N \sum_{\sigma, \sigma^\prime} v_{pqrs}^\text{eff} a_{p \sigma}^\dagger a_{q \sigma^\prime}^\dagger a_{r \sigma^\prime} a_{s \sigma},
\label{eq:H_eff}
\end{align}
where the indices $p,q,r,s$ run over the selected localized states, $a$ and $a^\dagger$ are respectively the electron annihilation and creation operators, and $\sigma$ denotes the spin quantum numbers. The two-body matrix elements $v_{pqrs}^\text{eff}$ in~\cref{eq:H_eff} are defined as
\begin{align}
&v_{pqrs}^\text{eff}=[W^\text{E}]_{pqrs} \nonumber \\
& := \int d\bm{r}_1 d\bm{r}_2 \phi_p^*(\bm{r}_1) \phi_q^*(\bm{r}_2) W^\text{E}(\bm{r}_1, \bm{r}_2) \phi_r(\bm{r}_2) \phi_s(\bm{r}_1),
\label{eq:v_eff}
\end{align}
where $W^\text{E}$ describes the Coulomb interaction between the electrons in the quantum defect screened by the response of the polarizable environment~\cite{ma2021quantum, sheng2022green}. On the other hand, the one-body matrix elements $t^\text{eff}_{pq}$ are calculated as~\cite{sheng2022green},
\begin{equation}
t_{pq}^\text{eff} = H_{pq}^\text{KS} - t_{pq}^\text{dc},
\label{eq:t_eff}
\end{equation}
where $H_{pq}^\text{KS}$ denotes the matrix elements of the KS Hamiltonian, and $t_{pq}^d$ is the so-called double counting correction term whose expression can be exactly derived within the $G_0W_0$ approximation~\cite{sheng2022green}. This correction removes the contributions of the Hartree and exchange-correlation potentials, included in the DFT calculations of the supercell, from the effective Hamiltonian of the defect that will be simulated using the quantum algorithms.\par

The steps detailed in this section are illustrated in \cref{fig:hero}(a).

\subsection{The optically-active defect states}\label{ssec:dipole}

Simulating the optical properties of the quantum defect requires access to certain eigenvalues $E_k$ and eigenstates $\ket{E_k}$ of the effective Hamiltonian. The optically-active states of the defect are identified by computing the dipole transition amplitude between the ground $\ket{E_0}$ and the excited $\ket{E_i}$ states
\begin{equation}
|\bm{D}_{i0}|^2=|\langle E_i |\bm{D}| E_0 \rangle|^2,
\label{eq:d_amplitude}
\end{equation}
where $\bm{D}$ is the electric dipole moment operator~\cite{dipole-note},
\begin{equation}
\bm{D} = -\sum_{i=1}^{N_e} \bm{r}_i.
\label{eq:dipole1}
\end{equation}
In~\cref{eq:dipole1} $N_e$ denotes the number of active electrons occupying the localized states in the defect, and $\bm{r}$ the position operator. In second quantization, the dipole observable in~\cref{eq:dipole1} is written as
\begin{equation}
\bm{D} = \sum_{p,q=1}^N \sum_\sigma \bm{d}_{pq} a_{p\sigma}^\dagger a_{q\sigma},
\label{eq:dipole}
\end{equation}
where $\bm{d}_{pq}$ is the matrix element
\begin{equation}
\bm{d}_{pq} = -\int \phi_p^*(\bm{r}) \bm{r} \phi_q(\bm{r}) d\bm{r}.
\label{eq:d_me}
\end{equation}

The position of the main peaks in the absorption or emission spectra of the quantum defect are determined by the energy of the excited states with the largest dipole transition amplitude $|\bm{D}_{i0}|^2$. These amplitudes can also be used to compute the radiative decay rates and lifetimes of the defect excited states. The rate (in atomic units) of spontaneous emission from the $i$th excited state to the ground state of the defect is calculated as~\cite{dzuba2019calculations}
\begin{equation}
\gamma_{i} = \frac{4}{3} (\alpha w_{i0})^3 |\bD_{i0}|^2,
\label{eq:rad_rate}
\end{equation}
where $\alpha$ is the fine structure constant and $w_{i0}:=\Delta E_{i0} = E_i-E_0$ is the excitation energy. Accordingly, the lifetime of the excited states is given by the inverse of the rate $\tau_i = 1/\gamma_{i}$. We show in \cref{sec:qlago} how to estimate both the excitation energy and the dipole amplitude on a quantum computer with guaranteed precision.

\section{Quantum algorithms}\label{sec:qlago}

This section explains the quantum algorithms developed to compute the optically-active excited states of the defect $\ket{E_i}$ and energies $E_i$ up to chemical accuracy, as determined by their dipole transition amplitudes $|\bD_{i0}|^2 = |\braket{E_i|\bD|E_0}|^2$. A brief review of the circuits and cost of the quantum algorithm subroutines used throughout this work may be found in~\cref{appsec:preliminaries}. A schematic showing the output of our algorithms is show in Fig. \ref{fig:hero}(b).

The basic idea is the application of quantum phase estimation (QPE) on the state $\bm D \ket{E_0}$. Since $\bD\ket{E_0}$ can be expanded in the Hamiltonian basis states $\ket{E_i}$, the output of the QPE circuit allows us to sample the eigenvalues $E_i$ of $H_\text{eff}$ with probability distribution given by the dipole transition amplitude $|\bD_{i0}|^2$. We propose two approaches depending on the difficulty to compute a good approximation to the ground state. The first solution, termed the ``loaded state" approach, leverages classical methods to prepare $\bD\ket{E_0}$ while potentially sacrificing accuracy. The second method performs all calculations on the quantum computer, thereby referred to as the ``prepared state'' approach. The key routine used in both methods is the simulation of the time evolution dictated by the Hamiltonian. In particular, Hamiltonian simulation is the core subroutine of quantum phase estimation (QPE), used to measure the energy and project onto $\ket{E_i}\bra{E_i}$. 

Both algorithms require the same number of qubits, as determined by QPE, but the prepared state approach has a higher gate cost. On the other hand, the loaded state approach suffers from a potentially uncontrolled error on the $\bD\ket{E_0}$ approximation. Therefore, the loaded state algorithm is recommended for quantum defects with strongly correlated excited states but weakly correlated ground states, where $\ket{E_0}$ can be classically estimated accurately. We now discuss each approach with more detail, leaving a complete description and resource estimation to \cref{appsec:PL_line_quantum}.

\subsection{The loaded state approach}\label{sssec:loaded_init_state}
The steps to compute $|\bD_{i0}|^2$ are shown in \cref{alg:dipole_active}. The algorithm also computes $E_i$ and $\ket{E_i}$ up to chemical accuracy. 

\begin{algorithm}[H]
\caption{Quantum algorithm to compute excited states, excitation energies, and dipole transition amplitudes.
\label{alg:dipole_active}
}
\begin{algorithmic}[1]
        \State Prepare $\ket{E_0}$ on the classical computer for chosen values of the total spin $S$ and total spin projection $M_S$, e.g. $S=1$ and $M_S\in\{0,\pm1\}$. Typically, this state is expressed as a sum of Slater determinants~\cite{fomichev2023initial}.
        \State Compute $\bD\ket{E_0}$ on the classical computer.
        \State Use the Sum-of-Slaters (SoS) algorithm \cite{fomichev2023initial} to efficiently load $\bD\ket{E_0}$ onto the quantum computer. 
        \State Apply $\QPE(H_\eff)$ on $\bD\ket{E_0}$.
        \State Measure the energy register and repeat steps (3) and (4) $K$ times, letting $K_i$ be the number of instances having measured $E_i$.
        \State Output $K_i/K$ as the estimate for $|\bD_{i0}|^2/\|\bD\ket{E_0}\|^2$.
\end{algorithmic}
\end{algorithm}
We make a few remarks on the implementation of each step and the cost. The first step requires the preparation of the ground state in a given spin sector. This avoids mixing states with different spin quantum numbers, which allows us to resolve different eigenstates when applying QPE with chemical accuracy. 
For simplicity, we assume that excitation energies are separated by more than chemical accuracy $\epsilon$ within a given sector ($S,M_S$) for total spin $S$ and total spin projection $M_S \in \{-S, -S+1, \ldots, S-1, S\}
$. In the rare case that two or more eigenstates have identical $(S, M_S)$ numbers and are $\epsilon$-close in energy, the algorithms introduced in this section still work correctly, with the degenerate spaces playing the role of single eigenstates. 

We note that a low-quality ground state $\tilde{\ket{E_0}}$ has still the potential to improve the estimation of matrix elements over classical methods. Indeed, the algorithm eventually computes $|\braket{E_i|\bD|\tilde{E_0}}|^2$ for a chemically accurate excited state $\ket{E_i}$, which can reduce the error over classical methods due to their additional error in estimating the excited state $\ket{E_i}$. 
Future work should analyze the ground state quality needed to estimate ODMR quantities of interest accurately; 
we refer the reader to Ref.~\cite{fomichev2023initial} for a related discussion.

The second step in~\cref{alg:dipole_active}, i.e. the computation of $\bD\ket{E_0}$ on the classical computer, only needs to be done once, because the result can be stored in memory, saving significant costs. However, it must still be reloaded on the quantum computer. To implement QPE($H_\eff$), we need to simulate the evolution of the Hamiltonian. Here, we use the double-rank factorization algorithm refined in Ref.~\cite{lee2021even}. The starting point of this algorithm is a linear combination of unitaries (LCU) decomposition $H_\eff = \sum_i c_i U_i$ with an associated LCU 1-norm $\lambda_{H_\eff} = \sum_i |c_i|$ that impacts the cost. This simulation is typically much more costly than the Sum-of-Slaters (SoS) protocol, which simply loads the initial state on the quantum computer. The total gate cost $C$ of the algorithm is obtained by
\begin{align}
    C = \kappa_i \cdot (C_{\QPE} + C_{\text{SoS}}),
\end{align}
where $C_{\QPE}$ is the cost of QPE($H_\eff$), $C_{\text{SoS}}$ is the cost of the SoS algorithm and costs $O(L\log(L))$ Toffoli gates, and $L$ is the number of Slater determinants representing $\bD\ket{E_0}$. Lastly, $\kappa_i$ is the number of samples needed to compute $|\bD_{i0}|^2$ with error $\error$, determined by the user.

More concretely, we need to compute the sampling probability $|\bD_{i0}|^2/\bD_0^2$, where $\bD_0:=\|\bD\ket{E_0}\|$, with error at most $\error \bD_0^{-2}$. As explained in the~\cref{ssec:dipole}, dipole transition amplitudes determine the radiative decay rate and lifetimes of the excited states, which typically span several orders of magnitude in the nanosecond (ns) time scale~\cite{gao2021radiative, liu2021temperature}. Thus, we set $\error$ such that $\tau$ is estimated within 1 ns. Following this assumption, we show in \cref{appssec:error_analysis_sampling} that $\kappa_i=O(F_i^2)$, for 
\begin{align}\label{eq:K_PL_hybrid}
   F_i= \bD_0/(w_{i0} |\bD_{i0}|)^3.
\end{align}
Using amplitude estimation, we can achieve $\kappa_i = O(F_i)$ which comes with a larger depth, as explained in \cref{appsec:Amplitude_estimation}. Using the asymptotic cost of the double-factorization algorithm~\cite{lee2021even}, the algorithm has asymptotic gate complexity
\begin{align}
    C &= \tilde{O}\left (\frac{\bD_0}{(w_{i0} |\bD_{i0}|)^3}(N^{3/2}\lambda_{H_\eff}/\epsilon+C_{\text{SoS}})\right) \nonumber\\
    &= \tilde{O}\left (\frac{\bD_0N^{3/2}\lambda_{H_\eff}}{(w_{i0} |\bD_{i0}|)^3\epsilon}\right) \label{eq:cost_hyb},
\end{align}
where $\lambda_{H_\eff}$ is the LCU 1-norm, and $\epsilon\sim 1.6\cdot 10^{-3}$ Hartree (chemical accuracy). In~\cref{eq:cost_hyb} we assumed that $C_{\text{SoS}} \ll C_\QPE$, which is the case in practice. As the SoS algorithm eventually liberates any used auxiliary qubits, the QPE subroutine determines the qubit complexity. This implies a qubit costing of $\tilde{O}(N^{3/2})$, with typically large constant overheads due to the qubit/gate trade-off by a key subroutine, the quantum read-only memory (QROM). QROMs play a pivotal role in the double-factorization algorithm simulating the evolution of the Hamiltonian. Lastly, we discuss the practicality of this algorithm in certain scenarios. 

This algorithm's cost would considerably increase for excited states with small dipole matrix elements since estimating dipole transition amplitudes would require many samples. However, the dipole transition amplitudes of the lowest-lying defect excited states show large values, as discussed in~\cref {sec:application}. In addition, due to the finite (chemical accuracy) resolution of the energies, one can expect a significant binning of the results in the higher energy eigenspace where the density of energy levels with small $\bm{D}_{i0}$ increases. On top of this binning, keeping the dipole amplitude fixed as a function of the ground state energy gap decreases the sampling complexity $\kappa_i$ and cost.

We also note that evaluating the cost in~\cref{eq:cost_hyb} requires prior information about the system, such as $w_{i0},|\bD_{i0}|$. This may seem problematic as we need the algorithm result to estimate its runtime. 
In practice, one would first check the maximum circuit depth implementable on the quantum computer, and compare it to $C_\QPE + C_{\text{SoS}}$ or $\tilde{\kappa_i}(C_\QPE + C_{\text{SoS}})$, for some estimate of $\kappa_i$ (for example using less accurate classical methods). Thus we either continually sample or implement the amplitude estimation (AE) approach with larger and larger depths, until some convergence criterion or a maximum runtime budget is reached. The same principle applies to other algorithms presented in this work.

\subsection{The prepared state approach}\label{sssec:Quantum_approach}

In the prepared state approach, we still start from an approximate ground state $\tilde{\ket{E_0}}$ uploaded on the quantum computer and refine it through coarse QPE~\cite{fomichev2023initial} or any other quantum-based method. The main assumption in this approach is that we have only a quantum access to a high-quality ground state approximation $\ket{E_0}$.

Therefore, to implement $\bD \ket{E_0}$ on the quantum computer, we block-encode the electric dipole operator (see~\cref{appsec:PL_line_quantum}). Block-encoding is a method allowing the transformation of states by non-unitary operators. The block-encoding of non-unitary operators comes with a success probability. In this case, the success probability being estimated is $\bD_0^2/\lambda_D^2$, where $\lambda_D$ is the LCU 1-norm of $\bD$. Given that the QPE sampling estimates $|\bD_{i0}|^2/\bD_0^2$, the product of these two estimations yield an estimate for  $|\bD_{i0}|^2/\lambda_D^2$. Hence, the resource estimation involves an iterative sampling procedure that yields an estimation error of $\error \lambda_D^{-2}$. As a consequence, the complexity is scaled by $\kappa_i = O(F_i^2)$ for $F_i=\lambda_D/(w_{i0} |\bD_{i0}|)^3$, or $O(F_i)$ if using the amplitude estimation method discussed in more detail in~\cref{appsec:Amplitude_estimation}. 

However, we adopt a cheaper alternative to this sampling method, where we first measure the block-encoding success or failure, and implement the last QPE if successful. The savings in cost is on the order of
\begin{align}\label{eq:K_pl_full}
    F_i = (|\bD_{i0}|+\bD_0)/(w_{i0} |\bD_{i0}|)^3,
\end{align}
compared to the simultaneous measurement approach where $F_i=\lambda_D/(w_{i0} |\bD_{i0}|)^3$. Since $\lambda_D \gg \bD_0$, the cost reduction could be significant as the sampling complexity is of the same order as in the loaded state approach as $|\bD_{i0}| < \bD_0$.

Overall, the cost of the prepared state approach is given by
\begin{align}\label{eq:easy_cost_eq}
C = \kappa_i \cdot (C_{\QPE} + C_{\bD}),
\end{align}
where $C_{\bD}$ is the cost of block-encoding the dipole operator, which is smaller than $C_{\QPE}$. Using the double factorization algorithm and block-encoding resource estimation of $\bD$ in \cref{appsec:PL_line_quantum}, we obtain the asymptotic scaling
\begin{align}
    C &= O\left(\frac{\bD_0}{(w_{i0} |\bD_{i0}|)^3} \big(C_\QPE + C_{\bD}\big)\right) \nonumber \\
    &= \tilde{O}\left(\frac{\bD_0}{(w_{i0}|\bD_{i0}|)^3}\big(\frac{N^{3/2}\lambda_{H_\eff}}{\epsilon} + N\big)\right).
\end{align}
As in the loaded state approach, the qubit complexity scales as $\tilde{O}(N^{3/2})$, determined by the QPE routine. The QPE and block-encoding costs can be computed in terms of the system size and coefficients of $H_\eff$ and $\bD$. However, the matrix element parameter $|\bD_{i0}|$ or excitation energies $w_{i0}$ are system-dependent and like $p_0$, cannot be approximated a priori. Thus, in practice we proceed as in the loaded state algorithm, iteratively sampling or implementing AE with larger depths, until convergence is achieved or budget exhaustion.

Lastly, we note one limitation of our resource estimation in this and the previous approach; we do not take into account the exact classical or quantum cost of the preparation of the ground state, as this is a hard problem for which there is no universally agreed efficient method and requires its own study \cite{fomichev2023initial}. Particularly for the prepared state approach, we mentioned beforehand that we simply assume quantum access to the ground state. This may be achieved through various means, e.g. adiabatically or through QPE, and in each case, there would be additional costs. For the example of QPE, the failure probability in the initial state preparation scales the algorithmic cost overhead by $1/p_0$, where $p_0= |\braket{E_0|\tilde{E_0}}|^2$ is the overlap of the approximate ground state $\tilde{\ket{E_0}}$ with $\ket{E_0}$. While one must be aware of this cost scaling, it is most often the case that they cannot be numerically evaluated beforehand. Hence, following the convention used in Ref.~\cite{babbush2018encoding}, we do not take this factor into account in our resource estimation.

\section{Application: a negative boron vacancy in a 2D hBN cluster}\label{sec:application}
Two-dimensional hexagonal boron nitride (hBN) is a wide bandgap material that can host a variety of spin defects~\cite{liu2022spin}. Recently, the optical properties of the negatively-charged boron vacancy ($\text{V}_\text{B}^-$) have been experimentally investigated for quantum sensing applications~\cite{mathur2022excited, gottscholl2021room}. The simplicity of this defect makes it an ideal use case to benchmark the quality of the embedding approach used to build the system Hamiltonian, and to estimate the resources required to implement our quantum algorithms.

The effective Hamiltonian $H_\text{eff}$ used for the quantum simulations is constructed with the QDET method, as implemented in the \texttt{WEST} code (version \texttt{v5.5})~\cite{govoni2015large, yu2022gpu}. To validate the embedding approach we carried out exact diagonalization of $H_{\text{eff}}$ and compared the excitation energies with previous results in the literature \cite{barcza2021dmrg}. In addition, we compute the electric dipole transition amplitudes between the ground and lowest-lying excited states and estimate their radiative decay rates and lifetimes.

\begin{figure}[t]
\centering
\includegraphics[width=0.9 \columnwidth]{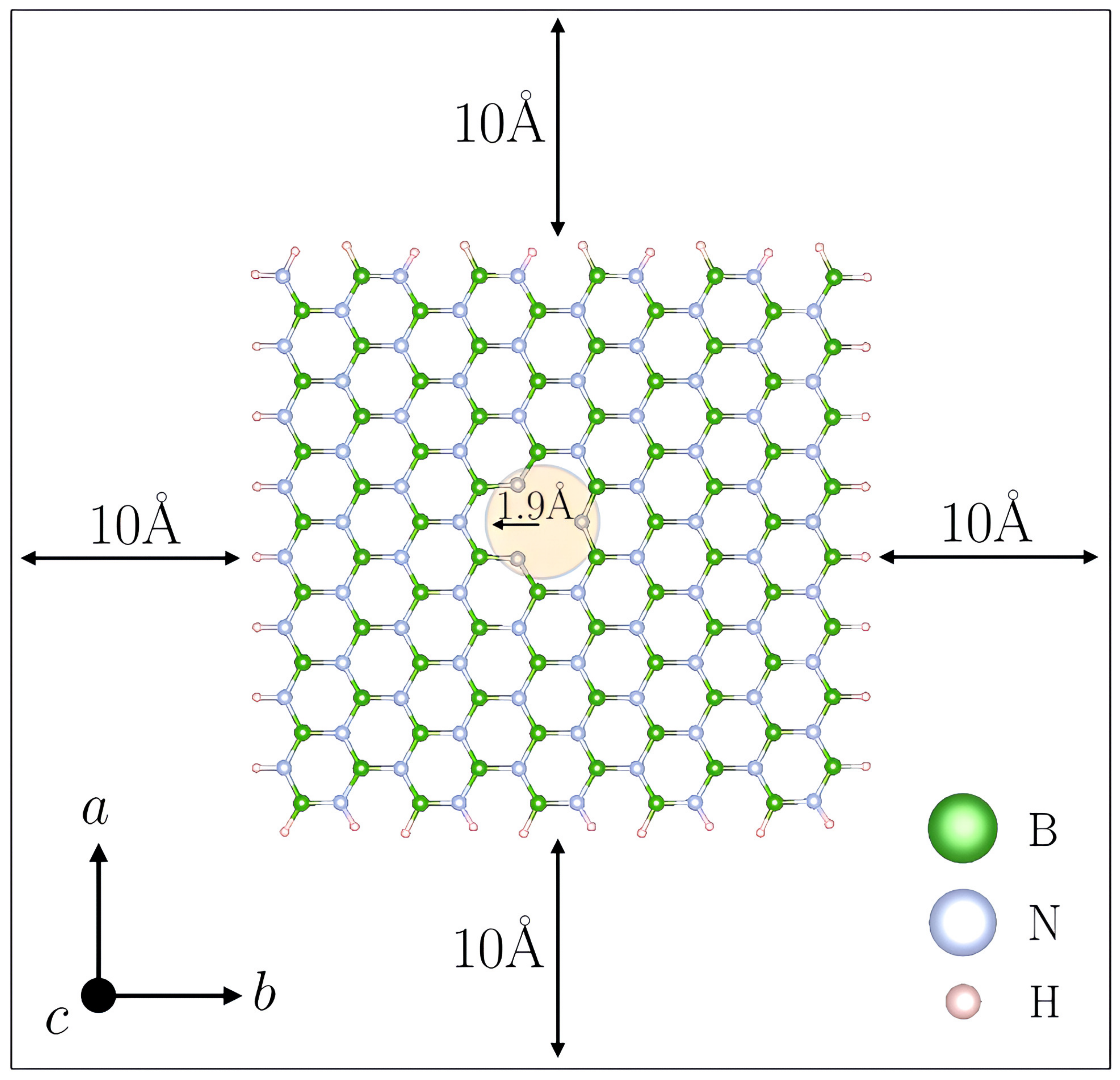}
\caption{Top view of the structural model used to simulate a negative boron vacancy $\text{V}_\text{B}^-$ in the center of a hexagonal boron nitride (hBN) cluster using an orthogonal supercell that includes  a vacuum layer of $10$ $\text{\AA}$ inserted in each direction. The cluster edges were passivated with hydrogen atoms. A sphere with radius 1.9 $\text{\AA}$ centered around the vacancy was used to compute the localization factor of the Kohn-Sham states to identify the defect localized states.}
\label{fig:structure}
\end{figure}

\cref{fig:structure} shows the supercell structural model of the defect-containing material. The negative boron vacancy was created in the center of a two-dimensional hexagonal boron nitride cluster consisting of 195 atoms with hydrogen-terminated edges ($\text{B}_{79} \text{N}_{80}\text{H}_{36}$). Within QDET, periodic boundary conditions are assumed to simulate the isolated defect in an extended material. We use an orthogonal supercell with primitive vectors $a=L(1, 0, 0), b=L(0, 1, 0)$, $c=L(0, 0, 0.5)$, where the lattice constant $L=40.25$ $\text{\AA}$ includes a vacuum layer of $10$ $\text{\AA}$ inserted in each direction to avoid spurious interaction between periodic images and account for the finite size of the cluster.

DFT calculations were performed using the \texttt{QUANTUM ESPRESSO} package(version \texttt{v7.2})~\cite{Giannozzi2020}. The core electrons of the atomic species were replaced with scalar relativistic norm-conserving 
pseudopotentials~\cite{Hamann2013}. The overall charge of the system was set to -1 to account for the charge state of the boron vacancy, and the PBESol~\cite{Perdew2008} density functional was used to estimate electronic exchange \& correlation effects. The optimized Kohn-Sham (KS) orbitals were represented in a plane-wave basis truncated using a kinetic energy cutoff of 75 Ry. With the defect center constrained to the $D3h$ symmetry, the cluster structure was relaxed within spin unrestricted DFT until ionic forces were less than $10^{-3}$ eV/\AA{}. The ground state occupations of the spin orbitals confirmed a triplet ground state at the level of DFT. Finally, the KS orbitals at the $\Gamma$ point were recomputed in the restricted open-shell framework with a band occupation fixed in the spin triplet configuration as to avoid spin contamination in the forthcoming many-body simulations. 

\begin{figure}[t]
\centering
\includegraphics[width=1 \columnwidth]{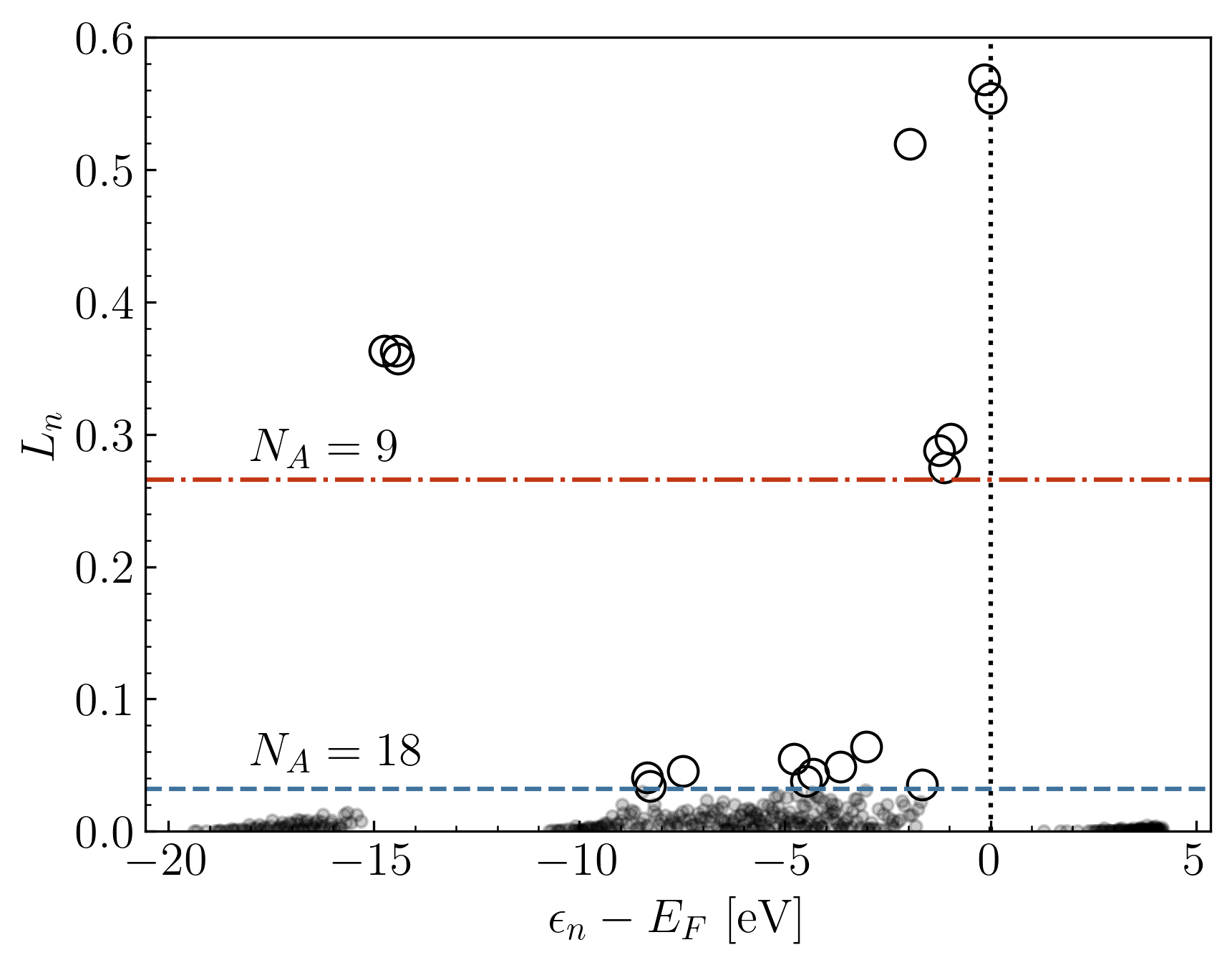}
\caption{Localization factor $L_n$ of the Kohn-Sham (KS) orbitals of the defect-containing hBN cluster. Small and faded black markers indicate states which are treated as delocalized environment orbitals while larger hollow black markers represent localized states in the active region at $N_A = 18$. KS energies $\epsilon_n$ are plotted relative to the Fermi energy $E_F$. The horizontal lines show that lowering the threshold value of $L_n$ increases the number of states in the active region from 9 to 18.}
\label{fig:localization}
\end{figure}

\cref{fig:localization} plots the localization factor $L_n$ (\cref{eq:loc_factor}) of the Kohn-Sham states. The integration volume $\text{V}_D$ used to compute $L_n$ was defined by a sphere with radius 1.9 $\text{\AA}$ centered at the position of the boron vacancy. This region encompasses the three nearest nitrogen sites to the vacancy. We observe nine strongly localized states with $L_n > 0.2$. Three of them are deep energy states appearing at $\sim 14$ eV below the Fermi energy ($E_F$), while the other six states emerge in the energy gap of the cluster. By gradually decreasing the threshold value of $L_n$ we can increase the number of single-particle states used to construct the effective Hamiltonian of the quantum defect.

We have built Hamiltonians of increasing sizes by considering active spaces with $N=6, 9, 11, 16$ and $18$ KS orbitals. The occupation of the KS states determines the number of electrons for each case. For this system, the two orbitals closest to the Fermi level are singly occupied while the other lower-energy states are doubly occupied. From~\cref{fig:localization} we see that decreasing the threshold localization factor adds doubly-occupied orbitals in the valence band. For example, the largest set of localized states corresponds to an active space with 34 electrons and 18 KS orbitals (34, 18). 

\begin{figure}[t]
\centering
\includegraphics[width=1 \columnwidth]{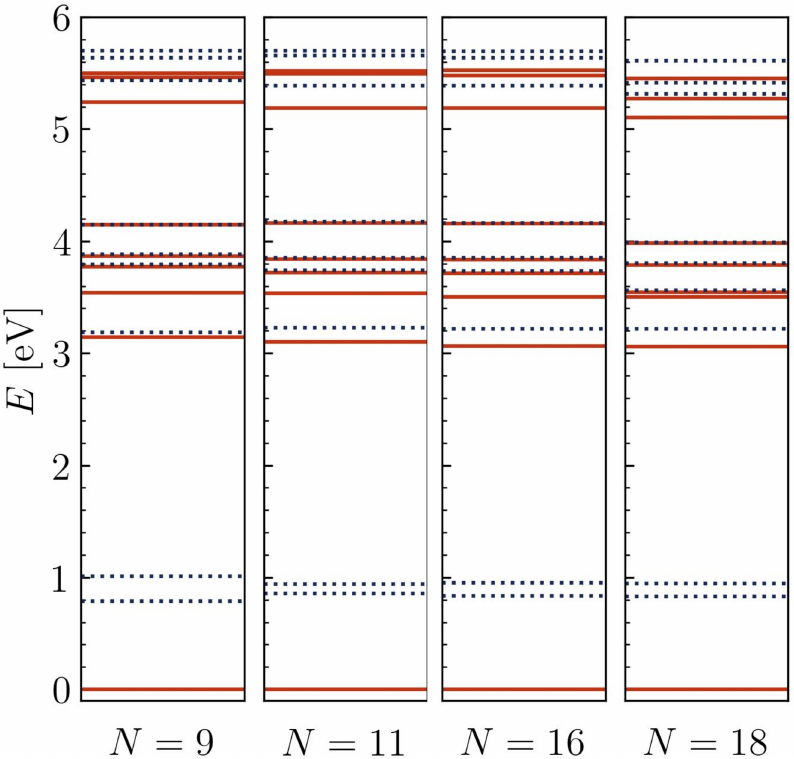}
\caption{Spectrum of electronic excitations of the quantum defect as the number of localized states $N$ used to compute the effective Hamiltonian is increased. Excitation energies were computed at the level of FCI in different sectors of the total spin with $S=1$ and $S=0$. Triplet and singlet states are distinguished in the plot by using solid (red) and dotted (blue) lines, respectively.}
\label{fig:spectra}
\end{figure}

\cref{fig:spectra} shows the spectra of vertical electronic excitations of the defect computed by using the full configuration interaction (FCI) method~\cite{jensen2017}. The ground and excited states were calculated using \texttt{PySCF}'s implementation of FCI as enabled using the \texttt{WESTpy}~\cite{westpy,sun2020recent} interface. The eigenstates with total spin $S=1$ and $S=0$ are reported in the \cref{fig:spectra} using solid and dotted lines, respectively. For all cases we found that the ground state of the defect is a triplet state. We observe small variations in the energies of the states with $E<5$ eV as we include states less strongly localized around the defect. The first excited states are singlet states with energies $0.8327$ eV and $0.9487$ eV relative to the triplet ground state. A qualitatively similar result is reported for a smaller cluster simulated with the density-matrix renormalization group (DMRG) method~\cite{barcza2021dmrg}. Higher-energy states exhibit a different structure as compared with the results reported in Ref.~\cite{barcza2021dmrg}. However, a direct comparison is not possible since we are simulating a larger system, and more importantly, we are including the effects of dielectric screening due to the host material via the embedding method.

\begin{table}[t]
\caption{Excitation energies $\Delta E_{i0}=E_i-E_0$, electric dipole transition amplitudes $|\bm{D}_{i0}|^2$, and estimated lifetimes of the defect excited states. $E_0$ is the lowest-energy state within each sector of the total spin $S$.}
\label{table:diptimes}
\centering 
\begin{tabular}{@{}>{\centering\arraybackslash}p{0.2\columnwidth}>{\centering\arraybackslash}p{0.26\columnwidth}>{\centering\arraybackslash}p{0.26\columnwidth}>{\centering\arraybackslash}p{0.26\columnwidth}@{}} 
\toprule 
& $\Delta E_{i0}$ (eV) & $|\bm{D}_{i0}|^2$ (a.u.) & $\tau_i$ (s) \\ 
\midrule 
 & $3.0586$ & $2.4914$ & $5.29 \times 10^{-9}$ \\
& $3.4999$ & $2.4487$ & $3.66 \times 10^{-9}$ \\
& $3.5467$ & $0.0314$ & $2.14 \times 10^{-5}$ \\
$S=1$ & $3.7889$ & $0.0280$ & $2.20 \times 10^{-5}$ \\
& $3.9820$ & $0.0833$ & $2.14 \times 10^{-6}$ \\
& $5.1002$ & $0.0083$ & $1.04 \times 10^{-4}$ \\
& $5.2692$ & $0.0385$ & $4.33 \times 10^{-6}$ \\
& $5.4493$ & $0.0227$ & $1.13 \times 10^{-5}$ \\
\midrule 
 & $0.1160$ & $1.8726$ & $1.72 \times 10^{-4}$ \\
& $2.3852$ & $2.1241$ & $1.54 \times 10^{-8}$ \\
& $2.7287$ & $0.0231$ & $8.65 \times 10^{-5}$ \\
$S=0$ & $2.9724$ & $0.0661$ & $8.20 \times 10^{-6}$ \\
& $3.1547$ & $0.0425$ & $1.66 \times 10^{-5}$ \\
& $4.4798$ & $0.0003$ & $1.58 \times 10^{-1}$ \\
& $4.5825$ & $0.0409$ & $5.82 \times 10^{-6}$ \\
& $4.7794$ & $0.0621$ & $2.23 \times 10^{-6}$ \\
\bottomrule 
\end{tabular}
\end{table}

We have used the eigenstates of the effective Hamiltonian for the largest active space (34, 18), to calculate the electric dipole transition amplitudes $|D_{i0}|^2$ using~\cref{eq:dipole}. This allows us to identify the optically-active states within each sector of the total spin ($S=1, 0$), and to estimate the radiative lifetimes of the lowest-lying excited states of the defect. The numerical results are reported in~\cref{table:diptimes}. The triplet states with the largest dipole amplitudes indicate spin-conserving transitions with energies of $3.06$ eV and $3.5$ eV. On the other hand, the brightest state with $S=0$ is obtained at $2.38$ eV. These excitation energies shed light on the position of the luminescence peaks associated with different de-excitation pathways via the triplet or the singlet channel. These predictions overestimate the position of 800nm (1.55 eV) $V_B^{-}$ luminescence peak \cite{mathur2022excited}. More accurate predictions would require optimizing the geometry for the excited states and computing the zero-phonon line~\cite{west-tddft}. Furthermore, because of the dependence of the DFT exchange and correlation functional on the quality of $H_{eff}$, higher rung functionals are likely to provide better results. Finally, the estimated radiative lifetimes for these states are of the order of a few nanoseconds, which is in good agreement with reported values in the literature~\cite{gao2021radiative, liu2021temperature}.

\begin{table*}[t] 
\centering 
\resizebox{\textwidth}{!}{
\begin{tabular}{@{}ccccc|ccc|ccc|ccc@{}} 
\toprule
\multirow{3}{*}{$N$} & \multirow{3}{*}{Qubits} & \multicolumn{6}{c}{\textbf{Loaded state approach}} & \multicolumn{6}{c}{\textbf{Prepared state approach}} \\
\cmidrule(lr){3-8} \cmidrule(lr){9-14} & &  \multicolumn{3}{c|}{\textbf{Standard Sampling}} & \multicolumn{3}{c|}{\textbf{Amplitude Estimation}} &  \multicolumn{3}{c|}{\textbf{Standard Sampling}} & \multicolumn{3}{c}{\textbf{Amplitude Estimation}} \\
\cmidrule(lr){3-8} \cmidrule(lr){9-14}
& & Depth & Samples & Total & Depth & Samples & Total & Depth & Samples & Total & Depth & Samples & Total \\
\midrule
$9$ & $523$ & $1.42 \times 10^8$ & $35$ & $5.26 \times 10^9$ & $2.23 \times 10^9$ & $6$ & $1.79 \times 10^{10}$ & $1.42 \times 10^8$ & $76$ & $1.11 \times 10^{10}$ & $2.23 \times 10^9$ & $9$ & $2.46 \times 10^{10}$ \\
$11$ & $608$ & $2.39 \times 10^8$ & $116$ & $3.08 \times 10^{10}$ & $3.75 \times 10^9$ & $11$ & $6.00 \times 10^{10}$ & $2.39 \times 10^8$ & $205$ & $5.24 \times 10^{10}$ & $3.75 \times 10^9$ & $15$ & $7.50 \times 10^{10}$ \\
$16$ & $1465$ & $7.39 \times 10^8$ & $2880$ & $2.13 \times 10^{12}$ & $1.16 \times 10^{10}$ & $54$ & $6.62 \times 10^{11}$ & $7.39 \times 10^8$ & $3070$ & $2.31 \times 10^{12}$ & $1.16 \times 10^{10}$ & $56$ & $6.85 \times 10^{11}$ \\
$18$ & $1633$ & $1.12 \times 10^9$ & $4850$ & $5.45 \times 10^{12}$ & $1.76 \times 10^{10}$ & $70$ & $1.35 \times 10^{12}$ & $1.12 \times 10^9$ & $5610$ & $6.35 \times 10^{12}$ & $1.76 \times 10^{10}$ & $75$ & $1.44 \times 10^{12}$ \\
\bottomrule
\end{tabular}
}
\caption{Constant factor resource estimation to compute the lowest-lying triplet excited state with the largest dipole transition amplitude $|\bm{D}_{i0}|^2$ for the $\text{V}_\text{B}^-$ in the hBN cluster described in \cref{sec:application}, for an increasing number of localized states $N$. Total qubit and gate complexities are reported for the easiest case (lowest cost) of estimating $|\bm{D}_{i0}|^2$ within an error threshold $\epsilon$, determined such that the lifetime $\tau_{i0}$ is computed within 1 ns accuracy. Depth and samples report the maximum depth and samples for computing the contribution of each component to the dipole transition amplitude.}
\label{tab:res_est_hBN}
\end{table*}

\section{Resource estimation}\label{sec:resources}
With the cost estimations in previous sections and the more accurate constant factor resource estimation carried out in the appendices, we provide a close estimate of the resources used by our algorithms.

We report in \cref{tab:res_est_hBN} the number of logical qubits and Toffoli gates used to compute the excitation energy and the dipole transition amplitude $|\bD_{i0}|^2$ of the brightest excited state with $S=1$ for the $\text{V}_\text{B}^-$ defect in the hBN cluster described in~\cref{sec:application}. We estimate the costs for both the loaded and prepared state approaches, and for different sizes of the effective Hamiltonian.

In \cref{tab:res_est_hBN}, we consider two strategies: the standard sampling approach which has a smaller quantum circuit depth (mentioned under `Depth') but more repetitions (`Samples'), and the amplitude estimation (AE) which has a longer-depth circuits than the sampling approach, by a factor of about $5\pi$ (\cref{appsec:Amplitude_estimation}), but requiring quadratically less many samples. The costs mentioned so far correspond to the largest depth and sampling complexity across the three axes for the dipole operator. However, the total gate complexity (`Total') for both sampling and AE is computed by adding up the product of depth and sampling complexities for all three axes. Note that the qubit complexity (`Qubits') is constant across all approaches, but it is much larger than $N$, due to the large overhead caused by the QROM in the double-factorization algorithm. In contrast, the gate and sampling complexities generally increase with system size. Lastly, we remark that the `Depth' for both the loaded and prepared state approaches are the same up to the three largest digits in the table. This is expected as the computational depth is overwhelmingly determined by the QPE, and not the block-encoding circuit employed exclusively in the second approach.

Beyond the sampling overhead, the cost is dominated by the QPE($H_\eff$) cost, which has been studied extensively in the literature \cite{babbush2018encoding,berry2019qubitization}. This implies that one area of focus for future research should be more efficient sampling strategies. However, there are statistical limits to the sampling complexity, at least in the current framework, and the QPE is the subroutine that requires the largest computational volume (number of gates $\times$ number of qubits). Therefore, to obtain a practical algorithm most efforts should be focused on the latter.

Overall, the amplitude estimation approach does not reduce the total complexity enough to make our algorithm suitable for early fault-tolerant quantum computers. Indeed, given the small size of the investigated defect, one should expect higher costs for classically intractable systems. In summary, despite the use of embedding methods to significantly reduce the size of the target Hamiltonian, \cref{tab:res_est_hBN} shows that quantum algorithmic improvements are still needed to make quantum computers useful for simulating quantum defects of increasing complexity.

\section{Conclusions and outlook} \label{sec:conclusions}

In this work, we investigated the application of quantum algorithms to simulate optically-active excited state properties of quantum defects. To mitigate the significant computational demands of simulating the entire defect-containing supercell, we employed QDET to build compact embedded Hamiltonians that encapsulate the effective screened interactions of electrons proximal to defect sites. These Hamiltonians were then utilized within the QPE sampling framework to selectively probe excited states with the largest dipole transition amplitudes from the ground state.

Using our framework, we estimated the quantum resources required to treat the negatively-charged boron vacancy in hexagonal boron nitride. Although our estimates align with those of other QPE-based algorithms in quantum chemistry \cite{Ivanov2023, steudtner2023fault}, our findings indicate that existing algorithmic approaches need further refinement to adapt to early fault-tolerant platforms. This need arises primarily because, although QPE is a cornerstone routine, its implementation—especially when paired with qubitization and double rank factorization—incurs substantial overhead. Additionally, our AE QPE variant introduces a compounded cost due to its nested QPE subroutines ~\cite{steudtner2023fault}. Future efforts should aim at significant reductions in computational overhead, potentially through minimizing the resources for a single QPE sample and decreasing the number of necessary samples by eliminating or suppressing the higher energy dipole transitions which are not of interest. Another avenue would be to find an alternative to the QPE sampling strategy altogether. Despite these challenges, the present work shows the potential for quantum algorithms to simulate complex quantum defects that are hard to treat with classical methods. By integrating embedding techniques with quantum computation, we can achieve accurate simulations of quantum defects with tightly controlled algorithmic error.

Extensions to this work could include the treatment of other processes which are key to discovering new ODMR-active defects. This includes determination of the fine structure of the defect levels, as generated by a multitude of phenomena including spin-spin, spin-orbit, and hyperfine interactions, as well as those produced by external fields. With fine structure-resolved energy levels, ODMR frequencies may now be accurately predicted. Two more vital components that should be addressed are intersystem crossing and magnetic dipole coupling between spin sublevels. The former is responsible for providing a pathway which ultimately permits an ODMR contrast signal, while the latter influences the rate of electron spin resonance. Moreover, considering the role of phonons, including electron-phonon interactions and phonon-assisted transitions, is essential for a comprehensive understanding of ODMR activity.

Overall, our findings and proposed extensions could position quantum computers to help advance technologies reliant on quantum defects, such as quantum sensors, single-photon sources, quantum memories, and even spin qubit-based quantum computers themselves. Our method provides a framework for precisely evaluating defect performance in scenarios challenging for classical approaches. Particularly, defects containing transition metals or rare-earth elements, which are stable in host materials like silicon carbide \cite{Bosma2018, Wolfowicz2020, Tissot2021}, 2D transition metal dichalcogenides \cite{Lee2022, Tsai2022, Lin2016}, and various Yttrium-based compounds \cite{Zhong2019, Yang2023, Raha2020}, exhibit complex behaviors due to the presence of localized and strongly correlated \textit{d} and \textit{f} electrons. These defects are challenging to accurately simulate even when they involve just a single substitutional site. Exploring more intricate defect configurations, including those with multiple defect sites, adjacent vacancies, and interstitial defects, further exposes the limitations of classical computational methods, thereby underscoring the critical role of our approach in bridging these gaps.

\section{Acknowledgments} \label{sec:acknowledgments}
This research used resources of the National Energy Research
Scientific Computing Center, a DOE Office of Science User Facility
supported by the Office of Science of the U.S. Department of Energy
under Contract No. DE-AC02-05CH11231 using NERSC award NERSC DDR-ERCAP0029552. We also acknowledge valuable conversations pertaining to this work with Stepan Fomichev.

\bibliographystyle{apsrev}
\bibliography{refs}

\begin{thebibliography}{62}
\bibitem[{dip()}]{dipole-note}
\bibinfo{note}{It is known that the position operator, and by extension, the dipole operator written as in Eq. (6) is ill-defined in periodic boundary conditions. While periodic boundary conditions are formally used in our simulations, our choice to work with an isolated system in a large supercell recovers the finite system limit where Eq. (6-8) hold.}
\end{thebibliography}

\onecolumngrid
\appendix

\section{Background on quantum algorithms}\label{appsec:preliminaries}

Here, we briefly review the subroutines used in the quantum algorithms and discuss their cost. We start with state preparation subroutines. A QROM is one such cornerstone subroutine. For a state represented by $d$ qubits with amplitudes given up to $b$ bits accuracy, the preparation costs $O(\lceil 2^d/x\rceil + bx)$ T-gates~\cite{low2018trading}, where $x$ represents an arbitrary power of 2 chosen to minimize the cost. QROM is based on the combination of SELECT and SWAP subroutines, and allows trading T-gates with qubits that need not be initialized. T-gates, along with Toffoli gates, belong to the class of non-Clifford gates, which cannot be transversally and fault-tolerantly implemented in most topological error codes, and consequently carry most of the implementation cost. Another important state preparation subroutine is known as Sum-of-Slaters~\cite{fomichev2023initial}. As the name suggests, it is tailored to prepare classically computed states represented with a small number $L$ of Slater determinants, each of which may require many qubits. It leverages QROM, and has a Toffoli cost of $(2\log L + 3)L$.

We also need to probabilistically implement non-unitary operators, achievable with a technique known as block-encoding~\cite{low2019hamiltonian}. Specifically, given a non-unitary operator $O$, block-encoding means finding states $\ket{G}$ and unitary operators $U$ such that
\begin{equation}\label{eq:block_encoding}
    O = (\bra{G}\otimes \bm{1})U(\ket{G}\otimes \bm{1}).
\end{equation}
In our case, we use block-encodings of the dipole operator $\bD$, the spin-spin Hamiltonian, $H_{SS}$, and the electronic Hamiltonian $H_\eff$. In all those cases, we implement the block-encoding by first decomposing the operator to a Linear Combination of Unitaries (LCU). The coefficients of such a linear combination become amplitudes in a superposition that are prepared in a PREP subroutine. After that, we apply the unitaries controlled on the basis states of said superposition, through a so-called SELECT subroutine. We finish the block encoding undoing the PREP, as suggested by~\cref{eq:block_encoding}. The result is a block-encoding, where $\ket{G} = \ket{0}$ in the superposition register, and $U$ represents the PREP, SELECT and PREP$^\dagger$ unitaries. 

Block-encoding and LCUs form the backbone of qubitization, one of the most popular techniques to implement Hamiltonian simulation~\cite{low2019hamiltonian}. This is, in turn, a core component of quantum phase estimation (QPE), a key algorithm. The latter approximates the transformation $\ket{\Psi} = \sum_j \alpha_j \ket{\bm{0}}\ket{E_j} \to \sum_j\alpha_j\ket{\text{Bit}(E_j)}\ket{E_j}$. In other words, it outputs the bitstring representation $\ket{\text{Bit}(E_j)}$ of energy $E_j$ of each eigenstate $\ket{E_j}$ in $\ket{\Psi}$. By sampling the energy register sufficiently many times, one can obtain the energy of a target eigenstate $\ket{E_j}$, and estimate the probability $|\alpha_j|^2 =|\braket{E_j|\Psi}|^2$. This method will allow us to estimate transitions of quantum operators. Since it is based on sampling, its cost scales as $O(\error^{-2})$ with the desired error $\varepsilon$. 
To reduce that scaling, one may use amplitude estimation, which combines quantum phase estimation with amplitude amplification to achieve a  $O(\error^{-1})$ scaling, at the expense of a larger depth by the same factor~\cite{brassard2002quantum} (see~\cref{appsec:Amplitude_estimation}).

As mentioned, to implement QPE for a Hamiltonian $H$, one needs to simulate the evolution operator $e^{-iHt}$. This is performed using qubitization~\cite{low2019hamiltonian}, which builds upon the block-encoding technique. The latter has many variants, and we use a state-of-the-art block-encoding called `double rank factorization'~\cite{von2021quantum,lee2021even}. Double rank factorization writes the Hamiltonian as a sum of quadratic terms, subsequently diagonalized via basis rotations. Among its advantages, it achieves a gate cost close to $\tilde{O}(N^{1.5})$, with a similar number of qubits. QPE requires $O(\lambda_{H}/\epsilon)$ controlled applications of the block-encoding, scaling the subroutine cost to $\tilde{O}(N^{1.5}\lambda_H/\epsilon)$. Here, $\lambda_H$ is the sum of the coefficients in the LCU, also called the LCU 1-norm, and $\epsilon$ is the tolerated error, often set at chemical accuracy $\sim 1.6\cdot 10^{-3}$ Hartree for our applications.

Put together, the block-encoding of $\bD$, quantum phase estimation of $H_\eff$ and the sampling of the energy register is the pipeline we employ to prepare $\ket{E_j}$, compute $E_j$, and estimate $|\braket{E_j | \bD | E_0}|^2$.

\section{Quantum algorithm for dipole transitions}\label{appsec:PL_line_quantum}
Here, we discuss the block-encoding of $\bD$ in full details along with the resource estimation.

\subsection{Overview: Block-encoding of dipole $D$}\label{appssec:block_encdoing dipole}
Assume $\frac{N_S}{2} = N$ spatial orbitals indexed by $p,q$ and $N_S$ total spin orbitals indexed with $p\sigma$, where $\sigma \in \{0,1\}$. Let
\begin{align}
    D := \sum_{p,q,\sigma} d_{pq} a_{p\sigma}a_{q\sigma}^\dagger 
\end{align}
be the dipole operator on one of the axes $x,y,$ or $z$. Note that the algorithm must be implemented for each axis. Diagonalizing yields
\begin{align}
    D = \sum_{k \in [N_S],\sigma} \lambda_k^D n_{k,\sigma}
\end{align}
where $n_{k,\sigma} = b_{k\sigma}b_{k\sigma}^\dagger$ and $b_{k\sigma}  = \sum_p \beta_{kp\sigma} a_{p\sigma}$. Following \cite{von2021quantum} (see also \cite[Sec. II.D]{lee2021even}), the number operator can be expressed through a series of Givens rotations starting at the first register, as follows: 
\begin{align}
    n_{k\sigma}  = U_{k\sigma}^\dagger n_{1\sigma}U_{k\sigma},
\end{align}
where $U_{k\sigma}$ is given by $\frac{N_S}{2}$ Givens rotations \cite[Lem. 8]{von2021quantum}. Recall $n_{1\sigma} = (\id - Z_{1\sigma})/2$ which leads to the following LCU
\begin{align}
    D = \tr(D) \id + \sum_{k,\sigma} \frac{\lambda_k^D}{2} U_{k\sigma}^\dagger (-Z_{1\sigma}) U_{k\sigma}
\end{align}
with the associated 1-norm $\lambda_D = \tr(D) + \sum_{k,\sigma} \frac{|\lambda_k^D|}{2} = \tr(D)+\sum_k |\lambda_k^D|$, as $\sigma \in \{0,1\}$. That being the case, we instead choose a less efficient LCU for now, which helps make the PREP state preparation process simpler 
\begin{align}
    D = \sum_{k,\sigma} \frac{\lambda_k^D}{2} \id  +  \frac{\lambda_k^D}{2} U_{k\sigma}^\dagger (-Z_{1\sigma}) U_{k\sigma}
\end{align}
with $\lambda_D = 2\sum_k |\lambda_k^D|$.

We now compute the costs of each subroutine. Notice there are no reflections. Only the cost of PREP and its inverse, along with SELECT. We aim to implement
\begin{align}
    \bra{\bm{0}} \PREP^\dagger \cdot \SEL \cdot \PREP \ket{\bm{0}} = \frac{D}{\lambda_D}
\end{align}

\subsection{PREP}
We need to prepare the state
\begin{align}
    \ket{+} \sum_{p,\sigma} \sqrt{\frac{|\lambda_p^D|}{2\lambda_D}} \ket{\theta_p}\ket{p\sigma}
\end{align}
where $\theta_p$ is the sign qubit for $\lambda_p^D$, and we note the change of variable from $k \to p$.

\subsubsection{Gate cost}
We shall compute the prepare and unprepare cost together. We start with preparing the nontrivial superposition over $\ket{p}$, meaning $\sum_{p,\sigma} \sqrt{\frac{2|\lambda_p^D|}{\lambda_D}} \ket{p}$,  which is done using advanced QROM (QROAM) and coherent alias sampling. Following QROAM most efficient clean ancilla implementations from \cite[App. B \& C]{berry2019qubitization}, the Toffoli cost for the QROAM involved in this preparation is 
\begin{align}
    \left\lceil \frac{N_S}{2k_1}\right\rceil + m(k_1-1)
\end{align}
and for the unpreparation is
\begin{align}
    \left\lceil \frac{N_S}{2k_2}\right\rceil + k_2
\end{align}
where $1 < k_1,k_2<N_S/2$ are chosen powers of two (if $k_1,k_2=1$, we have the schemes in \cite[Fig. 7]{babbush2018encoding} with Toffoli and ancilla cost $N_S/2-1, \lceil \log_2(N_S/2) \rceil$) and $m$ is the output size computed as
\begin{align}
 m=n_{N_S}+\mathfrak{N}+1   
\end{align}
where $n_{N_S}:=\lceil \log(\frac{N_S}{2}) \rceil$ is the number of bits used for the alt values and $\mathfrak{N}$ is the number of bits used for the keep values (determining accuracy) and one is for $\ket{\theta_p}$. We take $\mathfrak{N} \sim \lceil \log(1/\error)\rceil$ where $\error$ is the desired target error. Notice there is no $\lambda_D$ in this expression, as we are simply attempting to succeed at one application of block-encoding and the target error is not for the un-normalized but normalized $D/\lambda_D$. There are less significant costs, including 
\begin{itemize}
    \item Preparing the uniform superposition over $N_S$ basis states $\ket{\theta_p}\ket{p\sigma}$ for coherent alias sampling, requiring $3(n_N+1) - 3v_2(N_S) + 2b_r - 9$ Toffolis,  where $v_2(N_S)$ is the largest power of two factoring $N_S$, and $b_r$ is a number of bits used for rotation of an ancilla qubit to improve the amplitude of success. This cost is multiplied by two for unpreparation.
    \item The inequality test on the keep register, consuming $2\mathfrak{N}$ Toffolis in total for preparation and reverse. 
    \item The controlled SWAP done on the result of this inequality test is $2n_{N_S}$ for computation and uncomputation. Notice we do not swap the sign register as we can use the result of the inequality test to apply the required $Z$ gate. 
    \item Finally, the introduction of the first and last register can be simply done using two Hadamard gates.
\end{itemize}
In total, the Toffoli cost of PREP and its uncompute are
\begin{align}\label{eq:prep_gate_cost}
    \boxed{\left\lceil \frac{N_S}{2k_1}\right\rceil + m(k_1-1) +  \left\lceil \frac{N_S}{2k_2}\right\rceil + k_2 + 2(3(n_{N_S}+1) - 3v_2(N_S) 
    + 2b_r - 9) + 2\mathfrak{N} + 2n_{N_S}}
\end{align}
\subsubsection{Qubit cost}
For the qubit costings:
\begin{itemize}
    \item $(1+n_{N_S}+2) + (n_{N_S}+1) + \mathfrak{N} + \mathfrak{N} + 1 $ for the first $\ket{+}$ register and index, alt, keep, the register compared to keep, and the inequality test result registers, respectively.
    \item QROAM needs $m(k_1-1) + \lceil \log_2(N_S/(2k_1))\rceil $ ancillas which are rezeroed, and $k_2+\lceil \log_2(N_S/(2k_1))\rceil$ for the uncomputation which are also rezeroed. We only need the maximum of these two:
    \begin{align}
        Q_{\PREP} = \max(m(k_1-1) + \lceil \log_2(N_S/(2k_1))\rceil ,k_2+\lceil \log_2(N_S/(2k_1))\rceil)
    \end{align}
    \item There is one qubit needed for the rotations involved in the uniform superposition over $N$ basis states.
    \item The phase gradient state using $b_r$ qubits.
\end{itemize}
In total
\begin{align}\label{eq:prep_qubit_cost}
    \boxed{(3+n_{N_S}) + (n_{N_S}+1) + \mathfrak{N} + \mathfrak{N} + 1  + Q_{\PREP}  + 1 + b_r}
\end{align}

\subsection{SELECT\label{app:SELECT_D}}
The SELECT operation involves the following steps:
\begin{enumerate}
    \item Using QROAM to output the Givens rotations angles by reading $\ket{p\sigma}$. There are $\frac{N_S}{2}$ rotation angles, each given with accuracy $\mathfrak{M}$, giving an output $\otimes_{j=1}^{N_S/2}\ket{\theta_j^{(p\sigma)}}$  with size $\frac{N_S\mathfrak{M}}{2}$. The accuracy is determined later in \cref{sec:error_estimation}.
    \item The implementation of said Givens rotations on their respective registers $\ket{\cdot}_{(j-1)\sigma}\ket{\cdot}_{j\sigma}$. Notice the dependence of the registers on $\sigma$. Therefore, we make a CSWAP of the two contiguous registers controlled on $\sigma \in \{0,1\}$ to a working register and then apply the rotations. Notice we start at $j=N/2$ and we need to CSWAP back and forth to the working register for each new $j$.
    \item Once having reached $j=1$, controlled on the first PREP register $\ket{+}$, apply the $Z_{1\sigma}$ gate on the second working register. This is a simple CZ gate.
    \item Apply the inverse of the Givens rotations on the working register while C-SWAPping back and forth from the system into the working register.
    \item Apply the inverse of QROAM.
    \item Finally, apply the $Z$ gate on $\ket{\theta_p}$.
\end{enumerate}
\subsubsection{Gate cost}
Each step has the following costs. The Toffoli cost for the QROAM is
\begin{align}
    \left\lceil \frac{N_S}{k_1'} \right\rceil + \frac{N_S\mathfrak{M}}{2}(k_1'-1)
\end{align}
with uncomputation cost
\begin{align}
    \left\lceil \frac{N_S}{k_2'} \right\rceil + k_2' .
\end{align}
The implementation of the Givens rotations has Toffoli cost $N_S(\mathfrak{M}-2)/2$ with the same Toffoli cost for uncomputation, thus with total 
\begin{align}
    N_S(\mathfrak{M}-2).
\end{align}
Each CSWAP requires one Toffoli gate and each time we need to CSWAP two contiguous registers back and forth. We do this for all values of $j,\sigma$ and the whole process gets repeated to CSWAP back. Thus, we have in total 
\begin{align}
    2 ( N_S \cdot 2 \cdot 2) = 8N_S
\end{align} 
All the above brings the total SELECT Toffoli cost to
\begin{align}\label{eq:sel_gate_cost}
    \boxed{\left\lceil \frac{N_S}{k_1'} \right\rceil + \frac{N_S\mathfrak{M}}{2}(k_1'-1) + \left\lceil \frac{N_S}{k_2'} \right\rceil + k_2' + N_S(\mathfrak{M}-2) + 8N_S}
\end{align}

\subsubsection{Qubit cost}
The ancilla and output cost for the QROM is $\frac{N_S\mathfrak{M}}{2}(k_1'-1)+\lceil\log(N_S/k_1')\rceil, \frac{N_S\mathfrak{M}}{2}$ and the ancilla cost for its uncomputation is $\lceil \log(N_S/k_2')\rceil + k_2'$. Again, we only need the maximum of the computation and uncomputation ancilla costs, called 
\begin{align}
    Q_{\SEL} = \max\left(\frac{N_S\mathfrak{M}}{2}(k_1'-1)+\lceil\log(N_S/k_1')\rceil, k_2' + \lceil \log(N_S/k_2')\rceil\right)
\end{align}
The phase gradient state for the Givens rotations uses $\mathfrak{M}$ qubits. The system register qubit cost is $N_S$ and the working register is $2$. In total we have
\begin{align}
    \boxed{Q_{\SEL}+\frac{N_S\mathfrak{M}}{2} + N _S+ \mathfrak{M} + 2}
\end{align}

\subsection{Total costs}
\subsubsection{Gate cost}
This is simply the sum of \cref{eq:prep_gate_cost,eq:sel_gate_cost}. 

\subsubsection{Qubit cost}
Note that the QROAM ancillas in SELECT are rezeroed, similar to the QROAMs in the PREP part. We can therefore take a maximum between these ancilla costings. More precisely, replace $\frac{N_S\mathfrak{M}}{2}(k_1'-1)$ with 
\begin{align}
    Q_{\text{QROM}} = \max(Q_{\PREP}, Q_{\SEL})
\end{align}
Note that choosing even $k_1'=2$, is highly likely to make $Q_{\SEL}$ the maximum. The total qubit cost is
\begin{align}
    \boxed{[(3+n_{N_S}) + (n_{N_S}+1) + \mathfrak{N} + \mathfrak{N} + 1  +  1 + b_r] + [N_S+\mathfrak{M}+2] + Q_{\text{QROM}}}
\end{align}

\subsection{Error estimation}\label{sec:error_estimation}
There are three finite precision register size parameters. There is $b_r$, which is set to 7 to ensure a high success rate, so that its error has a negligible impact.
Then we have an approximation for the PREP state and an approximation on the SELECT for the Givens rotations. We have the following inequality
\begin{align}
    \left\| \sum_\ell \rho_\ell U_\ell - \sum_\ell \tilde{\rho_\ell} \tilde{U}_\ell\right\| \le \sum_\ell \rho_\ell \| U_\ell - \tilde{U}_\ell \|  + \sum_\ell |\rho_\ell - \tilde{\rho}_\ell| \le \error
\end{align}
here $\ell = (p,\sigma,x)$ where $x=0,1$ is used to denote identity or $Z_{1\sigma}$, respectively, $\rho_\ell = |\lambda_p^D|/(2\lambda_D)$, and $\sum \rho_\ell = \sum \tilde{\rho}_\ell = 1$. There are $L=2N_S$ many indices. For $x=0$, there is no approximation since SELECT is identity.

First let us treat the approximation by the Givens rotations. We know that each Givens rotation angle is approximated by $\mathfrak{M}$-bits, which implies that for each $\ell$ with $x=1$, we have $\| U_\ell - \tilde{U}_\ell \| \le  N_S \cdot \frac{\pi}{2^{\mathfrak{M}}}$, due to the total $N_S$ many Givens rotations (with their inverse), obtaining an overall SELECT error of
\begin{align}
    \left(\sum_{p,\sigma} \frac{|\lambda_p^D|}{2\lambda_D} \right) \frac{N_S}{2}\cdot \frac{\pi}{2^{\mathfrak{M}}} = \frac{1}{2}\cdot N_S\cdot \frac{\pi}{2^{\mathfrak{M}}}
\end{align}
For the PREP part error $\error_{\PREP} := \sum |\rho_\ell - \tilde{\rho}_\ell|$, let us denote by $\delta = \max_\ell |\rho_\ell - \tilde{\rho}_\ell|$. We know that $\error_{\PREP} \le \delta L = \delta (2N_S)$. So to achieve a desired target $\error_\PREP$, we can set a target of $\delta = \frac{\error_{\PREP}}{2N}$. By the coherent alias sampling process \cite[Eq. (35)]{babbush2018encoding}, we know 
\begin{align}
    \frac{1}{2^\mathfrak{N} L} \le \delta \implies \frac{2N_S}{\error_\PREP \cdot 2N_S} \le 2^\mathfrak{N}
\end{align}
and therefore, by simply picking $\mathfrak{N}= \lceil \log(1/\error_\PREP) \rceil$, we can achieve the desired PREP error.

The two PREP and SELECT errors should sum to less than desired $\error$, thus
\begin{align}
    \frac{1}{2^{\mathfrak{N}}} + \frac{N_S\pi}{2^{\mathfrak{M}+1}} \le \error
\end{align}
Since $\mathfrak{M}$ is involved in one of the most costly QROMs (outputting Givens rotations angles), we decide to concentrate 90\% of the accuracy on $\mathfrak{N}$, thus
\begin{align}
    \boxed{\mathfrak{N} = \left\lceil \log\left(\frac{10}{\error}\right) \right\rceil  , \ \ \mathfrak{M} = \left\lceil \log\left(\frac{9}{10}\cdot \frac{N_S\pi}{2\error}\right)\right\rceil}
\end{align}

\subsection{Resource estimation parameters}
In our resource estimation code, we choose 
\begin{itemize}
    \item $k_1 = 2^{\lfloor\log (\lceil(\sqrt{N_S/(2m)}\rceil) \rfloor}$
    \item $k_2 = 2^{\lfloor\log(\lceil(\sqrt{N/2}\rceil)\rfloor}$
    \item $k_1' = 1$
    \item $k_2' = 2^{\lfloor\log(\lceil(\sqrt{N_S}\rceil)\rfloor}$
    \item and target chemical accuracy $\error = 1.6e-3$.
\end{itemize}

\subsection{Error analysis of sampling}\label{appssec:error_analysis_sampling}
We analyze the prominent error sources in estimating the dipole transition amplitude from ground to excited states using our prepared state algorithm approach.

We assume access to an accurate ground state. We also assume that chemical accuracy used in the QPE can distinguish different energies in the lower eigenspace in each total spin sector $S$. Otherwise, the calculations below can be adapted by including a sum of $|\bD_{i0}|^2$ for $\ket{E_i}$ inside a QPE bin of length $\epsilon$.

There are two ways to perform the measurements in our computation. One way is to observe the success/failure of $\bD$'s block-encoding first, and if not failed, then continue with the QPE and the measurement on its energy register. The other is to both measurements at the same time at the end of the computation. Intuitively, the former should have the lower cost as we do not need to perform a QPE in the failed instances of the block-encoding. However, its error analysis is slightly more involved. For the sake of completeness, we show how to analyze both cases, but our resource estimation is only for the former approach.

\textit{Successive measurements ---}     We have a first error $\error_1$ in the estimation of $(\|\bD\ket{\psi}\|/\lambda_D)^2$ as a result of success/failure in the block-encoding.
Henceforth, we let $\bD_0 := \|\bD\ket{\psi}\|$.
The second error $\error_2$ is through our sampling after the QPE, in the estimation of $(|\bD_{i0}|/\bD_0)^2$. Through our sampling, we estimate the dipole transition amplitude by computing
\begin{align}
    \lambda_D^2 ((|\bD_{i0}|/\bD_0)^2 + \error_2) ((\bD_0/\lambda_D)^2+\error_1).
\end{align}
We would like the above to be $\error$ close to $|\bD_{i0}|^2$, where $\error$ is to be determined later. Taking the first order terms, the following should hold
\begin{align}\label{eq:error12_error}
    \lambda_D^2(|\bD_{i0}|/\bD_0)^2 \error_1 + \bD_0^2 \error_2 < \error,
\end{align}
where $\error$ shall be determined later. Notice that each $\error_i$ determine a sampling complexity scaling with $1/\error_i^2$.
The sampling complexity for estimating a probability $p$ with error $\error_0$ is $\frac{p(1-p)}{\error_0^2} < \frac{p}{\error_0^2}$. Given that these two samplings happen in a sequence, and the QPE measurement happens only after the success of the block-encoding, with probability $(\bD_0/\lambda_D)^2$, the number of samplings $\frac{|\bD_{i0}|^2}{\bD_0^2\error_2^2}$ done for the QPE satisfies:
\begin{align}
    \frac{\bD_0^2}{\lambda_D^2} \cdot \frac{\bD_0^2}{\lambda_D^2\error_1^2} = \frac{|\bD_{i0}|^2}{\bD_0^2\error_2^2} \implies \error_2 = \frac{\lambda_D^2|\bD_{i0}|}{\bD_0^3}\error_1
\end{align}
Hence the block-encoding success probability estimation error $\error_1$ must satisfy
\begin{align}
    \error_1 < \error \cdot \Big(\frac{\lambda_D^2 |\bD_{i0}|^2 + \lambda_D^2|\bD_{i0}|\bD_0}{\bD_0^2}\Big)^{-1}.
\end{align}
This yields
\begin{align}
    C_\text{Total} \le  \Big(\frac{|\bD_{i0}|(|\bD_{i0}| + \bD_0)}{\error}\Big)^2 \cdot (C_\text{QPE} + C_\text{Block-Encode(D)}) + \Big(\frac{\lambda_D|\bD_{i0}|(|\bD_{i0}| + \bD_0)}{\error}\Big)^2C_\text{Block-Encode(D)},
\end{align}
where the first term estimates the cost of when block-encoding is successful and followed by the QPE calculations, with sampling complexity $\frac{\bD_0^2}{\lambda_D^2} \cdot \frac{\bD_0^2}{\lambda_D^2\error_1^2}$, and the second where it fails, with sampling complexity upper bounded by $\frac{\bD_0^2}{\lambda_D^2\error_1^2}$. Asymptotically speaking, we observe that one can safely drop the last term as even with the $\lambda_D$ scaling, due to $C_\text{QPE} \gg C_\text{Block-Encode(D)}$, the first term is by far larger than the second. Lastly, we recall that the total complexity scales linearly if using amplitude estimation.

\textit{Simultaneous measurement ---} As we perform the QPE, the probability we estimate is $(|\bD_{i0}|/\bD_0)^2  \cdot (\bD_0/\lambda_D)^2 = |\bD_{i0}|^2/\lambda_D^2$. The question is how many samples we need to achieve a later to be determined error $\error$ in estimating $|\bD_{i0}|^2$. Formally,
\begin{align}\label{eq:error_definition_sampling}
    | \lambda_D^2 ( |\bD_{i0}|^2/\lambda_D^2 + \error_0) - |\bD_{i0}|^2 | \le \error \implies \error_0 = \error \lambda_D^{-2}
\end{align}
Recall that the sampling complexity for estimating a probability $p$ with error $\error_0$ is $\frac{p(1-p)}{\error_0^2} < \frac{p}{\error_0^2}$; this assumes that we do not use amplitude estimation technique which scales as $1/\error_i$ but has higher depth and constant factors.

So for $p=\frac{|\bD_{i0}|^2}{\lambda_D^2}$, we get the following cost estimate of our algorithm
\begin{align}\label{eq:cost_sampling}
    C_\text{Total} \le   \left(\frac{|\bD_{i0}|\lambda_D}{\error}\right)^2 \cdot (C_\text{QPE} + C_\text{Block-Encode(D)}),
\end{align}
where we have computed the block-encoding cost of $D$ computed in the previous sections, and the QPE cost is the cost of performing QPE on $H$ with chemical accuracy using the double-factorization algorithm.

\textit{Choosing $\error$ ---} Now let us determine $\error$. We set a target of $1\text{ns}$ error in the lifetime calculation, and compute the necessary $\error$. Accounting for all the units, the lifetime $\tau_{i0}$ is related to $\bD_{i0}$ by~\cite{dzuba2019calculations}
\begin{align}
    \tau_{i0} = \frac{3A}{4(\alpha w_{i0})^3 |\bD_{i0}|^2},
    \end{align}
where $A = 2.4189 \times 10^{-17},\alpha = \frac{1}{137}$ are constants and $w_{i0}= E_i-E_0$ is the gap between the excited and ground state. Targeting a $10^{-9}$ error in this approximation implies
\begin{align}
    \left|\frac{3A}{4(\alpha w_{i0})^3 (|\bD_{i0}|^2\pm \error)}-\frac{3A}{4(\alpha w_{i0})^3 |\bD_{i0}|^2}\right| < 10^{-9}
\end{align}
which simplifies to
\begin{align}
    \error < 10^{-9}\frac{4(\alpha w_{i0})^3 \cdot ||\bD_{i0}|^2\pm \error|\cdot |\bD_{i0}|^2 }{3A} \sim  10^{-9}\frac{4(\alpha w_{i0})^3 |\bD_{i0}|^4}{3A}
\end{align}
where the last estimation is reasonable as $\error$ is unlikely to be on the same scale as $|\bD_{i0}|$.
Substituting the values for $A,\alpha$,
\begin{align}
    \error = \frac{10^{-9} \cdot 4 \cdot 3.889 \cdot 10^{-7}\cdot w_{i0}^3 |\bD_{i0}|^4}{3 \cdot 2.4189 \cdot 10^{-17}} = 2.149\cdot 10\cdot w_{i0}^3 |\bD_{i0}|^4.
\end{align}
\textit{Total Cost ---} The total cost for the first measurement approach is estimated as
\begin{align}
    C_{\text{Total}} \le (\frac{|\bD_{i0}| + \bD_0}{2.149\cdot 10 \cdot w_{i0}^3 |\bD_{i0}|^3})^2  \cdot (C_\text{QPE} + C_\text{Block-Encode(D)}),
\end{align}
while for simultaneous measurement approach, it is estimated as
\begin{align}
     C_{\text{Total}} \le \left(\frac{\lambda_D}{ 2.149\cdot 10\cdot w_{i0}^3 |\bD_{i0}|^3}\right)^2 \cdot (C_\text{QPE} + C_\text{Block-Encode(D)}).
\end{align}
We emphasize that above equations apply when using the prepared state approach. Assuming classical access to $D\ket{\psi}$ and therefore $\bD_0$, we simply need to replace $\lambda_D$ by $\bD_0$ in the last cost formulae and drop the block-encoding cost, meaning
\begin{align}
    C_{\text{Total}} \le \left(\frac{\bD_0}{ 2.149\cdot 10\cdot w_{i0}^3 |\bD_{i0}|^3}\right)^2 \cdot C_\text{QPE}.
\end{align}

\section{Amplitude estimation\label{appsec:Amplitude_estimation}}
An alternative to sampling is to use amplitude estimation. As we saw in~\cref{eq:cost_sampling}, we want to estimate $p = \frac{|\bD_{i0}|^2}{\lambda_D^2}$ to error $\epsilon$. Such probability translates into an amplitude $a = \sqrt{p} = \frac{|\bD_{i0}|}{\lambda_D}$.
Then,~\cite[Theorem 12]{brassard2002quantum} indicates that for any $k\in\mathbf{N}$, the amplitude estimation algorithm with $M$ evaluations of the function returns an estimation $\tilde{a}=\sqrt{\tilde{p}}$ fulfilling
\begin{align}
     |\tilde{p}-p|\leq 2\pi k \frac{\sqrt{p(1-p)}}{M} + k^2 \frac{\pi^2}{M^2}\leq \varepsilon_0
\end{align}
with probability at least $8/\pi^2$ if $k = 1$ and $1-\frac{1}{2(k-1)}$ if $k\geq 2$. Choosing $k = 1$,
\begin{equation}
    M \geq \frac{2\pi k }{\varepsilon_0}\left(\sqrt{p(1-p)}+\frac{k\pi}{2M}\right)\geq \frac{2\pi k\sqrt{p} }{\varepsilon_0} = \frac{2\pi ka }{\varepsilon_0}.
\end{equation}
From~\cref{eq:error_definition_sampling},
\begin{equation}\label{eq:number_rep_AE}
    M \geq \frac{2\pi k|\bD_{i0}|\lambda_D^2 }{\varepsilon\lambda_D} = \frac{2\pi |\bD_{i0}|\lambda_D }{\varepsilon}.
\end{equation}
Similar equations apply for the loaded and prepared state solutions where we use the successive measurements approach.

Amplitude estimation on $U\ket{\phi}$ requires a walk operator $W= UR_\phi U^\dagger R_\phi^\dagger$ that includes the state preparation $R_\phi\ket{\bm{0}}=\ket{\phi}$, which in our case would be the SoS algorithm, and the $U$ itself, which would be potentially the block-encoding of $\bD$ followed by the QPE on $H_\eff$. Performing QPE on $W$ requires controlling the operations in $W$.

The QPE subroutine inside $U$ may be controlled by controlling just the initial Hadamard gates and the Hadamard gates in the QFT. Overall this amounts to twice as many Toffoli gates as the number of measurement qubits of the QPE, which is negligible compared to the actual cost of simulating the evolution. Similarly, the Sum-of-Slaters technique may be controlled by controlling the application of the first QROM, and in particular, its initial SELECT component, duplicating its cost. Finally, to control the block-encoding of $\bD$ and $H'_\text{SS}$, we only need to control the multi controlled-$Z$ gates in their SELECT component, detailed in \cref{app:SELECT_D} and Fig. 16 of~\cite{lee2021even}, respectively. In all instances of our approaches, given that $C_{\QPE(H_\eff)}$ overwhelmingly determines the overall cost, and given its occurrence twice in $W$ (once in $U$ and once in $U^\dagger$), we conclude that the factor $5\pi$ instead of $2\pi$ in \cref{eq:number_rep_AE} is a good estimate for an upper bound of the constant factor as a result of using the AE approach. Indeed, $2\times 2\pi$ to account for how many times $C_{\QPE(H_\eff)}$ is repeated, and one more $\pi$ factor to account for the control on the Hadamards, and the controlled SoS and $\bD$'s block-encoding algorithm. For example, \cref{eq:cost_sampling} becomes
\begin{align}
    C_\text{Total} \le   \Big(\frac{|\bD_{i0}|\lambda_D}{\error}\Big) \cdot (5\pi C_\text{QPE} + C_\text{Block-Encode(D)}).
\end{align}

\end{document}